\documentclass[12pt]{article}
\usepackage{amsfonts}
\renewcommand{\theequation}{\arabic{section}.\arabic{equation}}
\newcommand{\text}{\rm}

\begin{document}

\title{\textbf{On The Complete Seiberg-Witten Map For Theories With Topological Terms}}
\author{L. C. Q. Vilar, O.S. Ventura$^{a,b}$, R. L. P. G. Amaral$^{c}$, \\ V. E. R. Lemes$^{d}$  and L. O. Buffon$^{b,e}$ 
\footnote{mbschlee@terra.com.br, ozemar@cefetes.br, rubens@if.uff.br, vitor@dft.if.uerj.br, lobuffon@terra.com.br}\\
{ \small \em $^a$Coordenadoria de Matem\'{a}tica, Centro Federal de Educa\c {c}\~{a}o Tecnol\'{o}gica \  do Esp\'{\i}rito Santo},  \\
\small\em $ $Avenida Vit\'{o}ria 1729 - Jucutuquara, Vit\'{o}ria - ES, 29040 - 333, Brazil\\
\small\em $^b$Centro Universit\'{a}rio de Vila Velha, Rua comiss\'{a}rio Jos\'{e} \\
\small\em Dantas de Mello 15 - Boa Vista, Vila Velha - ES, 29102 - 770, Brazil\\
\small\em $^c$Instituto de F\'\i sica, Universidade Federal Fluminense, 24210 - 340, Niter\'oi - RJ, Brasil,\\
\small\em $^d$Instituto de F\'\i sica, Universidade do Estado do Rio de Janeiro,\\ 
\small\em Rua S\~{a}o Francisco Xavier 524, Maracan\~{a}, Rio de Janeiro - RJ, 20550-013, Brazil \\
\small\em $^e$Escola Superior de Ci\^{e}ncias da Santa Casa de Miseric\'{o}rdia de Vit\'{o}ria,\\ \small\em Av. Nossa Senhora da Penha 2190, Santa Luiza, Vit\'{o}ria-ES, 29045-402, Brazil. }
\bigskip
\maketitle

\vspace{-1cm}
\begin{abstract}
 The SW map problem is formulated and solved in the  BRST cohomological approach. The well known ambiguities of the SW map are shown to be associated to distinct cohomological classes. This analysis is applied to the noncommutative Chern-Simons action resulting in the emergence  of $\theta$-dependent terms in the commutative action which come from the nontrivial ambiguities. It is also shown how a specific cohomological class can be choosen in order to map the noncommutative Maxwell-Chern-Simons theory into the commutative one.
\end{abstract}
\setcounter{page}{0}\thispagestyle{empty}

\vfill\newpage\ \makeatother

\renewcommand{\theequation}{\thesection.\arabic{equation}}

\section{Introduction}

Distinct gauge choices in the open strings lead both to the realization of
ordinary Yang-Mills field theories as well as to noncommutative field
theories. It was the perception of this fact that made Seiberg and Witten
propose what became known as the Seiberg-Witten map (SW map) \cite{sw}. In
brief words, this map establishes a transformation of noncommutative field
variables in terms of ordinary (commutative) fields, in such a way that the
noncommutative gauge transformation is mapped into the ordinary one.

Let us express this mathematically. First we introduce the Moyal product
between two functions defined on the noncommutative space \cite{moyal}:

\begin{eqnarray}
f*g &=&\exp \left( \frac i2\theta ^{ij}\frac \partial {\partial x^i}\frac
\partial {\partial y^j}\right) f\left( x\right) g\left( y\right)
_{y\rightarrow x}  \nonumber  \label{bffgh} \\
&=&fg+\frac i2\theta ^{ij}\partial _if\partial _jg+O\left( \theta ^2\right) 
\text{ },  \label{moyal}
\end{eqnarray}
where the real c-number parameter $\theta ^{ij}$ comes from the
noncommutativity of the space-time coordinates

\begin{equation}
\left[ x^i,x^j\right] =i\theta ^{ij}.
\label{theta}
\end{equation} 
Making $\theta ^{ij}$ vanish brings the noncommutative theory into the
commutative one. Then, after defining the Moyal bracket,

\begin{equation}
\left[ f\stackrel{*}{,}g\right] =f*g-g*f\text{ },  \label{moyal com}
\end{equation}
we are able to construct a noncommutative gauge transformation,

\begin{equation}
\stackrel{\wedge }{\delta }_{\stackrel{\wedge }{\lambda }}\stackrel{\wedge }{%
A}\,=\partial _i\stackrel{\wedge }{\lambda }+\,i\left[ \stackrel{\wedge }{%
\lambda }\, \stackrel{*}{,}\, \stackrel{\wedge }{A_i}\right] =\,\stackrel{\wedge }{%
D}_i\stackrel{\wedge }{\lambda }\text{ },  \label{transf}
\end{equation}
where the hat symbol identifies fields and operators defined on the
noncommutative space-time and $\stackrel{\wedge }{D}_i$ represents the Moyal
covariant derivative. At this point it should be observed that, although
similar in form to a nonabelian gauge transformation, a nonvanishing
contribution coming from the Moyal bracket in (\ref{transf}) is expected even for an abelian gauge field. In the same way, the noncommutative gauge curvature

\begin{equation}
\stackrel{\wedge }{F}_{ij}=\partial _i\stackrel{\wedge }{A_j}-\partial _j%
\stackrel{\wedge }{A_i}-i\left[ \stackrel{\wedge }{A_i}\,\stackrel{*}{,}\,%
\stackrel{\wedge }{A_j}\right]  \label{fmini}
\end{equation}
gets a nonvanishing contribution coming from the commutator even for an
abelian field. In the abelian case, up to first
order in $\theta ^{ij}$, we get

\begin{equation}
\stackrel{\wedge }{\delta }_{\stackrel{\wedge }{\lambda }}\stackrel{\wedge }{%
A}\,=\partial _i\stackrel{\wedge }{\lambda }-\theta ^{kl}\partial _k%
\stackrel{\wedge }{\lambda }\partial _l\stackrel{\wedge }{A}_i+O\left( \theta ^2\right),
\label{delta a ex}
\end{equation}

\begin{equation}
\stackrel{\wedge }{F}_{ij}=\partial _i\stackrel{\wedge }{A_j}-\partial _j%
\stackrel{\wedge }{A_i}+\theta ^{kl}\partial _k\stackrel{\wedge }{A}_i\partial _l\stackrel{\wedge }{A}_j+O\left(
\theta ^2\right).  \label{fmini ex}
\end{equation}

Now, the sense of the SW map is that noncommutative gauge equivalent fields
should be mapped into ordinary gauge equivalent fields. If we express this
map as a formal series $\stackrel{\wedge }{A}_i\left( A\right) $, what we
are saying is that

\begin{equation}
\delta _\lambda \left[ \stackrel{\wedge }{A}_i\left( A\right) \right]
=\left[ \stackrel{\wedge }{\delta }_{\stackrel{\wedge }{\lambda }}\stackrel{%
\wedge }{A}_i\right] \left( A\right) \,,  \label{sw condi}
\end{equation}
where $\delta _\lambda $ is the ordinary gauge operation acting on $A_i.$
>From the abelian expression (\ref{delta a ex}), we see that the SW map cannot
be a simple field redefinition 
$\stackrel{\wedge}{A}_i=\stackrel{\wedge}{A}_{i}\hspace{-2 mm}(A,\theta)$ together with a reparametrization of $\stackrel{%
\wedge }{\lambda }=\stackrel{\wedge }{\lambda }\left( \lambda,\theta
\right) $. The $U\left( 1\right) $ case is symptomatic: such choice of
redefinitions would never get rid of the Moyal term in (\ref{delta a ex}),
which comes from the noncommutative nature of space-time, in contrast with
the absence of noncommutative terms in an ordinary $U\left( 1\right) $ gauge
transformation. The only hope is to mix $\lambda $ and $A$ in the SW map of $%
\stackrel{\wedge }{\lambda }$: $\stackrel{\wedge }{\lambda }=\stackrel{%
\wedge }{\lambda }\left( \lambda ,A,\theta \right) $.

In first order in $\theta $, we can write \cite{sw}

\begin{equation}
\stackrel{\wedge }{A}_i\left( A\right) =A_i-\theta ^{kl}\left( A_k\partial
_lA_i-\frac 12A_k\partial _iA_l\right) +O\left( \theta ^2\right) \text{ },
\label{asw}
\end{equation}

\begin{equation}
\stackrel{\wedge }{\lambda }\left( \lambda ,A\right) =\lambda +\frac
12\theta ^{kl}\left( \partial _k\lambda \right) A_l+O\left( \theta ^2\right)
\,.  \label{lsw}
\end{equation}
One can also establish how $\stackrel{\wedge }{A}_i$ and $\stackrel{\wedge }{%
\lambda }$ would change if we allow for variations in the $\theta ^{ij}$
parameter. This problem is, in fact, analogous to that of the SW map
in first order, as it is its solution \cite{sw}:

\begin{equation}
\delta _\theta \stackrel{\wedge }{A}_i=-\frac 14\delta \theta ^{kl}\left\{ 
\stackrel{\wedge }{A}_k\stackrel{*}{,}\partial _l\stackrel{\wedge }{A}_i+%
\stackrel{\wedge }{F}_{li}\right\} \text{ },  \label{3.8a}
\end{equation}

\begin{equation}
\delta _\theta \stackrel{\wedge }{\lambda }=\frac 14\delta \theta
^{kl}\left\{ \partial _k\stackrel{\wedge }{\lambda }\stackrel{*}{,}\stackrel{%
\wedge }{A}_l\right\} \,.  \label{3.8l}
\end{equation}

Equations (\ref{3.8a}) and (\ref{3.8l}) represent a system of coupled differential
equations. But, as we will discuss in sections $3$ and $4$, this system is not the most general one compatible with the SW condition (\ref{sw condi}). And, from this point of view, equations (\ref{asw}) and (\ref{lsw}) only 
represent a first order particular solution. This is the source for the ambiguity of the SW
map already observed in the literature \cite{sw}, \cite{asakawa}. It
also should be stressed that eq. (\ref{3.8a}) has been the starting point for
an interesting result, when it was argued that the SW map would transform
the 3D noncommutative Chern-Simons theory into the ordinary Chern-Simons
theory with no corrections in $\theta ^{ij}$ whatsoever. This first exact
result in the literature for the SW map was obtained by just showing that 
\cite{gs} 

\begin{equation}
\delta _\theta \stackrel{\wedge }{S}_{NCCS}=0\text{ }  \label{gs}
\end{equation}
where $\stackrel{\wedge }{S}_{NCCS}$ stands for the 3D noncommutative
Chern-Simons ($NCCS$) action and $\delta _\theta $ is the operation defined
in eq. (\ref{3.8a}). 

This result surprises as it is not obvious that the
substitution of equation (\ref{asw}) on the $NCCS$ action would promote an
exact cancellation of the $\theta$ terms order by order in $\theta$, 
in such a way that all the
noncommutative contributions coming from the Moyal products be gone, leaving
just the usual commutative Chern-Simons action,

\begin{equation}
\stackrel{\wedge }{S}_{NCCS}\stackrel{SW\,map}{\longrightarrow }S_{CS}\,.
\label{gss}
\end{equation}
As a comparison, no such result has been found up to now for any other
theory. Then, the questions are if one can find other examples of such
perfect mappings and what would be the property for a theory to allow these
mappings. Answering these questions is the aim of this paper.

So, we begin the work in the next section by exploring the ambiguity in the
solution of (\ref{3.8a}) and (\ref{3.8l}) from a different point of view, coming
from the cohomology of the operators in play. We will derive the complete solution of the cohomology problem associated to the SW map.  This will clarify the question if these ambiguities would damage
the exact SW map of $NCCS$. We will also briefly discuss the integrability of equation (\ref{3.8a}) in higher orders in $\theta $. With these developments, we will be able to generalize this equation and show
that also a theory as noncommutative Chern-Simons plus Maxwell ($NCMCS$) in 3D can be
exactly mapped into the ordinary $MCS$ theory avoiding any $\theta$ corrections. Our final conclusion claims for a conjecture that such exact
mappings will always be possible in the presence of topological terms, but
also that $\theta$ corrections will be unavoidable when such terms are
absent in pure geometrical theories. Some quantum aspects can then be
expected from this conjecture.

\section{Ambiguities}
\setcounter{equation}{0}
>From this point on, we will restrict ourselves to the abelian U(1) case.
The question of the ambiguities on the SW map can be traced from the disclosure
of the map itself. Soon it was noticed \cite{sw,asakawa} that the solution of the SW map (\ref{asw}) was not
uniquely defined. This can be foreseen from the definition of the map. It
only requests that noncommutative gauge equivalent classes should be mapped
into commutative gauge equivalent classes, so leading to a commutative gauge
invariant theory in the end. But this theory will remain gauge invariant if
we allow for a subsequent transformation in the commutative gauge field as
long as it has the form of a (field dependent) gauge transformation or a
gauge invariant field redefinition \cite{sw,asakawa}. So, if we make a new transformation made by the composition of the original SW map (\ref{asw}) and the gauge transformation or field redefinition of the commutative
connection, we will end with a map which will be again a SW map.

Some questions naturally arise in this context. The first is if there is
space for other ambiguities than those that we have listed above. The second
is what would be the (mathematical and physical) consequences of these
ambiguities in the SW mapping of a given noncommutative theory.

We will approach these questions from the point of view of BRST
cohomological techniques \cite{zumino,barnich,pqs}. In
this language, the gauge transformation of the commutative gauge field is described by
the action of the nilpotent BRST differential $s$,

\begin{equation}
sA_i =\partial _ic, \hspace{.3cm} \label{sasc} \medskip sc =0\text{ }, 
\hspace{.3cm} s^2 =0\text{ },  \nonumber
\end{equation}
where $c$ is the commutative ghost field. As before, we take $\stackrel{\wedge }{C}$ and$\stackrel{\wedge }{A}_i$, the noncommutative fields, as
formal power series in $\theta $ whose coefficients are local polynomials in 
$c$ and $A_i$ (and in their derivatives). Then equation (\ref{sw condi})
for the SW map is translated as \cite{zumino}

\begin{equation}
s\stackrel{\wedge }{A}_i=\partial _i\stackrel{\wedge }{C}+\text{ }i\left[ 
\stackrel{\wedge }{C}\stackrel{*}{,}\stackrel{\wedge }{A}_i\right] \text{ },
\label{sahat}
\end{equation}
and 
\begin{equation}
s\stackrel{\wedge }{C}=i\stackrel{\wedge }{C}*\stackrel{\wedge }{C}\,.
\label{schat}
\end{equation}
We write the series in $\theta $ as

\begin{equation}
\stackrel{\wedge }{A}_i=A_i+\sum_{n=1}^\infty A_i^{(n)},  \label{ahat}
\end{equation}
\begin{equation}
\stackrel{\wedge }{C}=c+\sum_{n=1}^\infty C^{(n)},  \label{chat}
\end{equation}
where $A_i^{(n)}$ and $C^{(n)}$ identify the term with $n$ $\theta ^{\prime
}$s in the power series expansion of the fields. Now we are able to expand
equations (\ref{sahat}) and (\ref{schat}) as well. In their $n^{\prime }$th order, we find 

\begin{eqnarray}
sA_i^{(n)}&=&\partial _iC^{(n)}-\sum_{\alpha =1}^n\theta ^{kl}\partial
_kC^{(n-\alpha )}\partial _lA_i^{(\alpha -1)}+.....\\
&+&\frac{i^{n+1}}{2^{n}n!}\theta
^{k_1l_1}...\theta ^{k_nl_n}\left( \partial _{k_1}..\partial _{k_n}c\partial
_{l_1}..\partial _{l_n}A_i - \partial _{k_1}..\partial _{k_n}A_i\partial
_{l_1}..\partial _{l_n}c \right) , \nonumber \label{san}
\end{eqnarray}
\begin{eqnarray}
\hspace{-5mm}sC^{(n)}&=&-\frac 12\sum_{\alpha =1}^n\theta ^{kl}\partial _kC^{(n-\alpha
)}\partial _lC^{(\alpha -1)}+... \\ &+& \frac{i^{n+1}}{2^{n}n!}\theta
^{k_1l_1}...\theta ^{k_nl_n}\partial _{k_1}..\partial _{k_n}c\partial
_{l_1}..\partial _{l_n}c\,, \nonumber \label{sac}
\end{eqnarray}
with $A_i^{(0)}=A_i$ and $C^{(0)}=c$. Then, up to second order in $\theta$,

\begin{equation}
sA_i=\partial _ic\text{ },  \label{sa0}
\end{equation}
\begin{equation}
sA_i^{(1)}=\partial _iC^{(1)}-\theta ^{kl}\partial _kc\partial _lA_i\text{ },
\label{sa1}
\end{equation}
\begin{equation}
sA_i^{(2)}=\partial _iC^{(2)}-\theta ^{kl}\left( \partial _kC^{(1)}\partial
_lA_i+\partial _kc\partial _lA_i^{(1)}\right) \text{ },  \label{sa2}
\end{equation}

and 
\begin{equation}
sc=0\text{ },  \label{sc0}
\end{equation}
\begin{equation}
sC^{(1)}=-\frac 12\theta ^{kl}\partial _kc\partial _lc\text{ },  \label{sc1}
\end{equation}
\begin{equation}
sC^{(2)}=-\theta ^{kl}\partial _kc\partial _lC^{(1)}\text{ }.  \label{sc2}
\end{equation}

The first order in $\theta $ equation, where the map begins, has a solution which can
be divided into two parts. The first part is any particular solution of the
``inhomogeneous'' equations, where the ``inhomogeneity'' is characterized by
the explicit $\theta $-dependent term in (\ref{sa1}) and (\ref{sc1}). Let
us call this part as $a_i^{(1)}$ and $c^{(1)}$ respectively. The second part
is the general solution $(\mathbb{A}^{(1)},\mathbb{C}^{(1)})$ for the associated
homogeneous equations

\begin{equation}
s\mathbb{A}_i^{(1)}=\partial _i\mathbb{C}^{(1)}\text{ },  \label{sah}
\end{equation}
and 
\begin{equation}
s\mathbb{C}^{(1)}=0\text{ }.  \label{sch}
\end{equation}

The first part of the solution, $\left( a_i^{(1)},c^{(1)}\right) $, became
known as the first order SW map \cite{sw}, which can be read
from eqs.(\ref{asw}) and (\ref{lsw})

\begin{eqnarray}
a_i^{(1)}(A) &=&-\theta ^{kl}\left( A_k\partial _lA_i-\frac 12A_k\partial
_iA_l\right) \text{ },  \label{a1} \\
c^{(1)}\left( c,A\right) &=&\frac 12\theta ^{kl}\left( \partial _kc\right)
A_l\text{ }.  \nonumber
\end{eqnarray}
Now, the general solution $\mathbb{A}_{i}^{(1)}$ is the source of all the
ambiguities and freedom in the SW map that the literature refers to \cite{sw,asakawa}. Eq.(\ref{sah}), together with the
nilpotency of $s$ assured in (\ref{sasc}), happens to be a problem of the
cohomology of the BRST operator modulo total derivatives (a review of the
cohomology of BRST can be found in \cite{livro}). $\mathbb{A}_{i}^{(1)}$ is the most general polynomial satisfying (\ref{sah}) constructed
with the commutative gauge field, derivatives and one $\theta $, in the
sector with ghost number zero, canonical dimension $1$ and carrying a free
Lorentz index (remembering eqs. (\ref{moyal}) and (\ref{theta}), we can associate a dimension $-2$ to $\theta $). Eq. (\ref{sch}), which is a direct consequence of the
nilpotency of $s$ and of eq. (\ref{sah}), is a problem of the local
cohomology of $s$, in the sector with ghost number $1$ and zero canonical
dimension.

In fact, the set of eqs.(\ref{sah}) and (\ref{sch}) will be found in each
sector $n$ of the $\theta $ expansion of (\ref{sahat}) and (\ref{schat})%
\begin{equation}
s\mathbb{A}_i^{(n)}=\partial _i\mathbb{C}^{(n)}\text{ },  \label{sahn}
\end{equation}
and 
\begin{equation}
s\mathbb{C}^{(n)}=0\text{ }.  \label{schn}
\end{equation}
The solutions will change as the required number of $\theta ^{\prime }$s
change for each $n$.

In the study of the cohomology of BRST modulo total derivatives ($\mathcal{H}$($s/d$)) one
can find an analogous set of equations as this above. They are known as
descent equations, and they appear in the analysis of the quantum stability
and anomalies of gauge theories \cite{livro}. In the seek of
completion, we will briefly review the general ideas of this study. Let us
write generic descent equations with the same Lorentz structure as (\ref
{sahn}) and (\ref{schn}) 
\begin{equation}
sW_i=\partial _iX\text{ },  \label{desc1}
\end{equation}
and 
\begin{equation}
sX=0\text{ },  \label{desc2}
\end{equation}
where $W_i$ and $X$ are formal power series in the field and parameter space
of the gauge theory. We have to solve this problem beggining by the last
equation of the set, (\ref{desc2}), which is a simpler problem of the local
cohomology. The solutions of (\ref{desc2}) are classified as trivial, those
which are written as simple BRST variations of polynomials in the field
space $P(\varphi )$ of the problem, or non-trivial solutions, invariant
polynomials which cannot be written in this way, 
\begin{equation}
X=Y+sZ\text{ },\text{ \qquad }\mid \text{ }sY=0\text{ }\hspace{1mm}and\hspace{1mm}\text{ }%
Y\neq s\hspace{1mm}\Omega \hspace{1mm},\text{ }\forall \hspace{1mm}\Omega \in P(\varphi )\text{ }.  \label{xsol}
\end{equation}
We say that $Y$ belongs to the local cohomology of the $s$ operator, $Y\in 
\mathcal{H}(s)$. We can act with a derivative on (\ref{xsol}) and
substitute the result on (\ref{desc1}), 
\begin{equation}
s\left( W_i-\partial _iZ\right) =\partial _iY\text{ }.  \label{wint}
\end{equation}
Let us say that we can find a particular solution to this equation, let us
call it $\delta (\partial _iY)$, such that 
\begin{equation}
s\delta (\partial _iY)=\partial _iY\text{ }.  \label{delta}
\end{equation}
It is important to observe that sometimes it is not possible to find such
particular solutions coming from non-trivial contributions of lower level
descent equations $\cite{livro}$. When this happens, we say
that this solution is obstructed and we are forced to make null its
coefficient in order to continue the procedure. Substituting (\ref{delta})
on (\ref{wint}), we get 
\begin{equation}
s\left( W_i-\delta (\partial _iY)-\partial _iZ\right) =0\text{ }.
\label{wlocal}
\end{equation}
Now, this is again a problem of the local cohomology of $s$, and again the
solutions will be classified as in (\ref{xsol})

\begin{equation}
W_i-\delta (\partial _iY)-\partial _iZ=Y_i+sZ_i\qquad \mid \text{ }%
sY_i=0\text{ }\hspace{1mm}and\hspace{1mm}\text{ }Y_i\neq s\Omega _i,\text{ }\forall \hspace{1mm}\Omega _i\in
P(\varphi )\text{ }.\label{cohow}
\end{equation} 
So, finally, the general solution of (\ref{desc1}) is 
\begin{equation}
W_i=Y_i+\delta (\partial _iY)+sZ_i+\partial _iZ\text{ }.  \label{w}
\end{equation}
The terms given by $Y_i+\delta (\partial _iY)$ represent the nontrivial
contributions to the cohomology of $s/d$, and $sZ_i+\partial _iZ$ is the
trivial part which is written as $s$ or $\partial _i$ variations of cocycles
in the field space. In many problems of interest in quantum field theory,
the trivial solutions are discarded for not carrying physical information.
This happens in the study of the stability and anomalies in QFT. In what
concerns the present problem of the SW map, this cohomological
classification will also be of mathematical and physical relevance, as we
will see in a moment.

We can return now to our specific problem. We want to solve the
cohomological problem posed by equations (\ref{sahn}) and (\ref{schn}) in
the field space generated by $A_i$, $c$, and derivatives acting on them, in
the presence of $\theta $ parameters (and obviously of all other parameters the
theory under study allows), 
\begin{equation}
P_{SW}=\left\{ \theta ,A_i,c,\partial _i\right\} \text{ }.  \label{fspace}
\end{equation}
We first argue that trivial $s$ cocycles do not exist at the level of the
upper descent equation (\ref{sahn}) for any $n$. Notice that $\mathbb{A}%
_i^{(n)}$ has zero ghost number, and as there is no element in the field
space $P_{SW}$ with negative ghost number, it is not possible to find any $%
Z_i$ with ghost number $-1$ as required by eq.(\ref{w}). Next we will prove
that neither there are contributions of the form $\delta (\partial _iY)$ at
the upper level .

\textbf{Proposition 2.1 }The only nontrivial solutions of the $s/d$
cohomological problem given by eqs.(\ref{sahn}) and (\ref{schn}) belong to
the local $s$ cohomology $\mathcal{H}(s)$ at the upper level descent
equation.

\textbf{Proof }: The local BRST cohomology of (abelian) gauge theories for
arbitrary quantum numbers has already been extensively studied. It is given
by polynomials built with the ghost $c$ non derivated and the curvature $%
F_{ij}=\partial _iA_j-\partial _jA_i$ possibly derivated \cite{livro}. In the case of eq.(\ref{schn}), we are looking for the
cohomology of $s$ in the sector of ghost number $1$, zero dimension and in
the presence of a $n$ number of $\theta ^{\prime }s$. So, the form of the
general element of $\mathcal{H}(s)$ in this sector is 
\begin{equation}
\alpha c\theta ^{k_1l_1}...\theta ^{k_nl_n}.P_{k_1l_1...k_nl_n}(\partial
_i,F_{ij})\text{ },  \label{hs}
\end{equation}
where $\alpha $ is a numerical coefficient and $P_{k_1l_1...k_nl_n}(\partial
_i,F_{ij})$ represent all possible $s$ invariant polynomials constructed
with an arbitrary number of derivatives of the curvature, with free indices $%
k_1l_1...k_nl_n$, and total dimension $2n$. This is the most general
nontrivial solution for $\mathbb{C}^{(n)}$. Following the steps from (\ref
{xsol}) to (\ref{delta}), in order to find the contribution in the upper
level coming from this lower level descent equation nontrivial solution, we
must solve 
\begin{eqnarray}
s\mathbb{A}_i^{(n)} &=&\alpha \partial _ic\theta ^{k_1l_1}...\theta
^{k_nl_n}.P_{k_1l_1...k_nl_n}(\partial ,F)+\alpha c\theta
^{k_1l_1}...\theta ^{k_nl_n}.\partial _iP_{k_1l_1...k_nl_n}(\partial ,F)%
\text{ } \nonumber \label{obstruct} \\
&=&s\left( \alpha A_i\theta ^{k_1l_1}...\theta
^{k_nl_n}.P_{k_1l_1...k_nl_n}(\partial ,F)\right) +
\nonumber \\ && +\alpha c\theta
^{k_1l_1}...\theta ^{k_nl_n}.\partial _iP_{k_1l_1...k_nl_n}(\partial ,F)%
\text{ }.  
\end{eqnarray}
It becomes clear now that there is no possible solution for $\mathbb{A}_i^{(n)}
$ as the last term on (\ref{obstruct}) is again an element of $\mathcal{H}%
(s)$ and in this way cannot be written as a $s$ variation of any cocycle of
the field space. This characterizes the kind of obstruction we just
mentioned, and the way out is to make $\alpha =0$. Then, the only
non-trivial solutions to (\ref{sahn}) are those of the local cohomology $%
\mathcal{H}(s)$ in the sector of ghost number zero, dimension equals to one,
in the presence of a $n$ number of $\theta ^{\prime }s$ and with a free
Lorentz index. \textit{\textbf{QED}}

Let us call this local cohomology at the upper level equation (\ref{sahn})
as $\mathcal{H}^{(n)}(s)$. The general element of $\mathcal{H}^{(n)}(s)$
will be of the form 

\begin{equation}
Y_i^{(n)}=\alpha \theta ^{k_1l_1}...\theta
^{k_nl_n}.P_{ik_1l_1...k_nl_n}(\partial ,F)\text{ },  \label{hns}
\end{equation}
where $\alpha $ is a numerical coefficient and $P_{ik_1l_1...k_nl_n}(%
\partial ,F)$ represent all possible $s$ invariant polynomials constructed
with an arbitrary number of derivatives of the curvature, with free indices $%
ik_1l_1...k_nl_n$, and total dimension $2n+1$. Finally, we find that the
most general solutions to the eqs.(\ref{sahn}) and (\ref{schn}) are given
by 
\begin{equation}
\mathbb{C}^{(n)}=sZ^{(n)}\text{ },  \label{cnsol}
\end{equation}
and 
\begin{equation}
\mathbb{A}_i^{(n)}=Y_i^{(n)}+\partial _iZ^{(n)}\text{ },  \label{ansol}
\end{equation}
where $Z^{(n)}$ are generic polynomials with zero ghost number and
dimension, built with elements of the field space $P_{SW}$ in the presence
of $n$ $\theta ^{\prime }s$. Notice that, unlike $Y_i^{(n)}$, the $Z^{(n)}$
polynomials need not be invariant under the action of the BRST operator.

Having solved the cohomology associated to the problem of the SW map, we can
now ask what changes the solutions $\mathbb{A}_i^{(n)}$ can bring in the higher
orders of the SW map once they are defined. For simplicity, let us say that
we have found these cocycles in a given theory already at the first order, $%
n=1$, 
\begin{eqnarray}
\mathbb{C}^{(1)} &=&sZ^{(1)}\text{ },  \label{a1cocy} \\
\mathbb{A}_i^{(1)} &=&Y_i^{(1)}+\partial _iZ^{(1)}\text{ }.
\end{eqnarray}
The second order equations for the SW map would be 
\begin{eqnarray}
sA_i^{(2)} &=&\partial _iC^{(2)}-\theta ^{kl}\left( \partial _k\left(
c^{(1)}+\mathbb{C}^{(1)}\right) \partial _lA_i+\partial _kc\partial _l\left(
a_i^{(1)}+\mathbb{A}_i^{(1)}\right) \right) \text{ },  \label{sa2cocy} \\
sC^{(2)} &=&-\theta ^{kl}\partial _kc\partial _l\left( c^{(1)}+\mathbb{C}%
^{(1)}\right) \text{ },
\end{eqnarray}
where $a_i^{(1)}$ and $c^{(1)}$ are the particular solutions given in 
(\ref{a1}). We can make our usual division of the solution into the particular
solution to the system above $\left( a_i^{(2)},c^{(2)}\right) $ and the
general solution of the associated homogeneous system $\left( \mathbb{A}%
_i^{(2)},\mathbb{C}^{(2)}\right) $. This latter system will be of the form of 
(\ref{sahn}) and (\ref{schn}) for $n=2$, and will be solved once we find
the polynomials $Y_i^{(2)}$ and $Z^{(2)}$, and substitute them on the
general equations (\ref{cnsol}) and (\ref{ansol}). The former system can
be further split into a part that depends on the solutions of the cohomology
of the first level, which we call $\left( a_i^{R(2)},c^{R(2)}\right) $, 
\begin{eqnarray}
sa_i^{R(2)} &=&\partial _ic^{R(2)}-\theta ^{kl}\left( \partial _k\mathbb{C}%
^{(1)}\partial _lA_i+\partial _kc\partial _l\mathbb{A}_i^{(1)}\right) \text{ },
\label{sa2redef} \\
sc^{R(2)} &=&-\theta ^{kl}\partial _kc\partial _l\mathbb{C}^{(1)}\text{ },
\end{eqnarray}
and a part $\left( a_i^{I(2)},c^{I(2)}\right) $ which only depends on the
particular solutions $\left( a_i^{(1)},c^{(1)}\right)$ of the first level, 
\begin{eqnarray}
sa_i^{I(2)} &=&\partial _ic^{I(2)}-\theta ^{kl}\left( \partial
_kc^{(1)}\partial _lA_i+\partial _kc\partial _la_i^{(1)}\right) \text{ },
\label{sa2indep} \\
sc^{I(2)} &=&-\theta ^{kl}\partial _kc\partial _lc^{(1)}\text{ }.
\end{eqnarray}
Different solutions of this last system have already been found in the
literature $\cite{okuyama},\cite{wess},\cite{fidanza}$. We
reproduce that of \cite{okuyama} for $a_i^{I(2)}$: 
\begin{equation}
a_i^{I(2)}=\frac 12\theta ^{kl}\theta ^{mn}A_k\left( \partial _lA_m\partial
_nA_i-\partial _lF_{mi}A_n+F_{lm}F_{ni}\right) \text{ .}  \label{ai2}
\end{equation}
But all these particular solutions that we cited are conected by trivial
cocycles of the form $\partial _iZ^{(1)}$ \cite{fidanza}, showing that they all belong to the same (trivial) cohomological class.
Then, the work to be done is to solve the system (\ref{sa2redef}). We
first notice that if it was not for the presence of the trivial cocycle $%
\mathbb{C}^{(1)}$, the first equation of (\ref{sa2redef}) would have exactly
the form of equation (\ref{sa1}) for $n=1$ with $\mathbb{A}_i^{(1)}$ placed
instead of $A_i$. So, our guess for the particular solution of 
(\ref{sa2redef}) is to take $a_i^{(1)}$ and $c^{(1)}$ and change $A_i$ for $\mathbb{A%
}_i^{(1)}$ in the following form 
\begin{eqnarray}
a_i^{R(2)} &=&-\theta ^{kl}\left( \mathbb{A}_k^{(1)}\partial _lA_i+A_k\partial
_l\mathbb{A}_i^{(1)}-\frac 12\mathbb{A}_k^{(1)}\partial _iA_l-\frac 12A_k\partial
_i\mathbb{A}_l^{(1)}\right) \text{ },  \label{a2redef} \\
c^{R(2)} &=&\frac 12\theta ^{kl}\left( \partial _kc\right) \mathbb{A}_l^{(1)}
\text{ }+ \frac{1}{2}\theta ^{kl}\partial_{k}\mathbb{C}^{(1)}A_{l}.  \nonumber
\end{eqnarray}
In fact, as can be inferred from equations (\ref{san}) and (\ref{sac}), the existence of a
solution $\mathbb{A}_i^{(n)}$, eq.(\ref{ansol}), at any given order $n$
developes a problem analogous to that of the first order upon the next order $n+1$, 
$\emph{i.e.}$%
\begin{eqnarray}
sa_i^{R(n+1)} &=&\partial _ic^{R(n+1)}-\theta ^{kl}\left( \partial _k\mathbb{C}%
^{(n)}\partial _lA_i+\partial _kc\partial _l\mathbb{A}_i^{(n)}\right) \text{ },
\label{sanredef} \\
sc^{R(n+1)} &=&-\theta ^{kl}\partial _kc\partial _l\mathbb{C}^{(n)}\text{ },
\end{eqnarray}
which will have the same kind of solution as (\ref{a2redef})
\begin{eqnarray}
a_i^{R(n+1)} &=&-\theta ^{kl}\hspace{-1mm}\left( \hspace{-1mm}\mathbb{A}_k^{(n)}\partial
_lA_i \hspace{-1mm}+\hspace{-1mm}A_k\partial _l\mathbb{A}_i^{(n)}\hspace{-1mm}-\hspace{-1mm}\frac 12\mathbb{A}_k^{(n)}\partial
_iA_l \hspace{-1mm}-\hspace{-1mm}\frac 12A_k\partial _i\mathbb{A}_l^{(n)}\hspace{-1mm}\right) \text{ },  \label{anredef}
\\
c^{R(n+1)} &=&\frac 12\theta ^{kl}\left( \partial _kc\right) \mathbb{A}_l^{(n)} + \frac{1}{2}\theta ^{kl}\partial_{k}\mathbb{C}^{(n)}A_l.  \nonumber
\end{eqnarray}

We can rewrite the expressions (\ref{a2redef}) in a compact way, making
it clear the implicit redefinition of $A_i$ and of $c$:
\begin{eqnarray}
a_i^{R(2)} &=&a_i^{(1)}\left( A_i+\mathbb{A}_i^{(1)}\right) \mid _{\theta ^2},
\label{anfinal} \\
c^{R(2)} &=&c^{(1)}\left( c + \mathbb{C}^{(1)},A_{i}+\mathbb{A}_i^{(1)}\right) \mid _{\theta ^2}, 
\nonumber
\end{eqnarray}
where $\mid _{\theta ^2}$means projection on the $\theta ^2$ dependence of
the polynomial. Obviously, this procedure can be continued to higher orders.
For example, it is not difficult to show that the effect of $\mathbb{A}_i^{(1)}$
on the third order particular solution of the SW map is $a_i^{(2)}\left( A_i+%
\mathbb{A}_i^{(1)}\right) \mid _{\theta ^3}$. And, as eqs.(\ref{anredef})
show, each new element $\mathbb{A}_i^{(n)}$ (\ref{ansol}) of the cohomology of 
$s$ modulo derivatives adjoined at each order in $\theta $ means a new
redefinition of the gauge conection. Recalling that the elements of $%
\mathcal{H}^{(n)}(s)$ are gauge invariant polynomials constructed with
curvatures and their derivatives, we indeed have shown the identification
between the BRST sense of the SW map ambiguity $\cite{pqs}\text{ , }
\cite{zumino} $, given by $\mathcal{H}^{(n)}(s)$, and the well known
freedom of the SW map by redefinitions of the commutative gauge potential 
$\cite{sw}\text{ , }\cite{asakawa}$. The BRST trivial part of
the solution (\ref{ansol}) for $\mathbb{A}_i^{(n)}$, being given by total
derivatives $\partial _iZ^{(n)}$, is easily seen as the freedom of the SW
map by field dependent gauge tranformations $\cite{sw}\text{ , }\cite{asakawa}$.

Finally, as there are no other possible elements allowed by the general
solution (\ref{ansol}), we answer the first question we proposed at the
beggining of this section by saying that there is no room for other
ambiguities in the SW map than these listed above.

It is time now to call attention to the different roles played by the
cohomologically trivial and non-trivial parts of the general solution of the
SW map. The trivial parts, as we just mentioned, have the form of gauge
transformations. So, if we sum a BRST trivial term to any particular
solution of the SW map, the final commutative gauge invariant action so
mapped will not change at all. But a nontrivial solution of $\mathcal{H}%
^{(n)}(s)$ in the cohomology of $s/d$ will be able to modify any commutative
action (see $\text{ }\cite{mapmcs}$ for a study of the effects
of gauge invariant field redefinitions on gauge invariant actions). After
the understanding of this point, it becomes clear that there can be no
uniquely defined commutative action coming from the SW map of a
noncommutative theory if $s/d$ nontrivial elements of $\mathcal{H}^{(n)}(s)$
are found. This is a first hint in the way to answer the second question at
the beggining of this section about the implications of the ambiguities on
the mapping of noncommutative actions. Anyway, in the next sections we will extract some general information on the commutative theories coming
from the SW map.


\section{The General SW Map for NCCS}
\setcounter{equation}{0}

>From the developments of the last section, we saw that eqs. (\ref{asw}) and 
(\ref{lsw}) are not a complete solution to the SW map problem. Already at
first order there is a contribution to the solution in eq. (\ref{asw}) coming
from elements of $\mathcal{H}(s)$ which will be relevant to the SW mapping
of noncommutative actions

\begin{equation}
\stackrel{\wedge }{A}_i(A)=A_i-\theta ^{kl}(A_k\partial _lA_i-\frac 12A_k\partial _iA_l)+%
\mathbb{A}_i^{(1)}+O(\theta ^2)\text{ .}  \label{32}
\end{equation}

We also saw how the presence of nontrivial contributions of a given order $n$
will alter the higher order terms, by generating a covariant mapping

\begin{equation}
A_i\rightarrow A_i+\mathbb{A}_i^{(n)}  \label{33}
\end{equation}

\noindent in the particular solution of the SW map.

With this in mind, we are now in a position to analyse eqs.(\ref{3.8a}) and 
(\ref{3.8l}) and understand the consequences of its use. As described before,
eq.(\ref{3.8a}) is a solution of a problem analogous to the first order SW
map, and thus it is subjected to the same limitations as those of the particular
solution (\ref{asw}). In fact, equation (\ref{3.8a}) is only valid modulo field
dependent gauge transformations and covariant field redefinitions 
(in \cite{asakawa} the authors have pointed out in this direction but incorporated only total
derivatives (trivial contributions) in their analysis). This
can also be noticed if one tries to write the second order expansion ($%
\theta \delta_\theta $) of (\ref{3.8a}) using as first order solution eq.(\ref
{asw}). Then, one finds that eq.(\ref{3.8a}) is not integrable at this order,
if such ``ambiguities'' are not taken into account. In a certain sense, eq.%
(\ref{3.8a}) only encodes information on the specific map associated to the
particular solution (\ref{asw}) of the SW map. Thus, results coming from the
use of eq.(\ref{3.8a}) will only refer to direct consequences of such a
particular map and will not allow for the broader picture implied by the
general solution (in the next section we will make use of the generalization of (\ref{3.8a})). 

This conclusion takes us back to eq.(\ref{gs}) developed in \cite{gs}. We said
that there can be no uniquely defined commutative action coming from the SW
map of a noncommutative theory when we have $s/d$ nontrivial $\mathcal{H}(s)$
contributions. But the conflict is now solved when we understand that eq.
(\ref{gs}) comes as a direct consequence of eq.(\ref{3.8a}), and the latter is, by
its turn, dependent on a particular solution of the SW map, the one
associated to the $s/d$ trivial solutions of $\mathcal{H}(s)$.

So, we will explicitly construct now a solution of the SW map with the
intent of showing how the commutative Chern-Simons theory can be deformed by 
$\theta $ terms coming from its noncommutative version.

In the first order in 3D, the only element of $\mathcal{H}(s)$ is

\begin{equation}
\mathbb{A}_i^{(1)}=\alpha \theta ^{kl}\partial _iF_{kl}  \label{34}
\end{equation}
(just remembering that the original $NCCS$ action does not contain the metric,
so we do not include it on the field and parameter content of the theory).
But this element becomes trivial in the $s/d$ cohomology as it is a total
derivative in the free index, and in this way it does not give any
contribution to the commutative action.

In the second order, we can find an element of $\mathcal{H}(s)$

\begin{equation}
\mathbb{A}_i^{(2)}=\alpha \theta ^{ab}\theta ^{ef}F_{ai}\partial _bF_{ef}
\label{35}
\end{equation}


\noindent which cannot be written as a total derivative in the free index.

This suggests a deformation of the commutative $CS$ action in the $\theta ^2$
order. In fact, writing the SW map up to second order as (compare with the
usual solution of \cite{okuyama} in (\ref{ai2}))

\begin{eqnarray}
\stackrel{\wedge }{A}_i &=&A_i-\frac 12\theta ^{kl}\left( 2A_k\partial
_lA_i-A_k\partial _iA_L\right)  \label{36} \\
&&+\theta ^{kl}\theta ^{mn}\left[ \frac 12A_k\left( (\partial _lA_m)\partial
_nA_i-(\partial _lF_{mi})A_n+F_{lm}F_{ni}\right) +\alpha F_{ki}\partial
_lF_{mn}\right] \nonumber \\ &&+o(\theta ^3)\text{ },\nonumber
\end{eqnarray}

\noindent we get a contribution from the SW map of the $NCCS$ action in the
commutative space beyond the usual $CS$ term,

\begin{eqnarray}
\stackrel{\wedge }{S}_{NCCS}\longrightarrow {} &&S_{CS}+\frac{\alpha}{2} \int d^3x\epsilon ^{abc}\theta
^{kl}\theta ^{mn}\times  \label{37} \\
&&\times\left[ F_{ka}\partial _lF_{mn}\partial _bA_c+A_a\partial _b\left(
F_{kc}\partial _lF_{mn}\right) \right] +o(\theta ^3)  \nonumber \\
&=&S_{CS}+\frac{\alpha}{2} \int d^3x\epsilon ^{abc}\theta ^{kl}\theta ^{mn}\left[
\partial _l\left( F_{ka}F_{mn}\right) F_{bc}\right] +o(\theta ^3)\text{ }\nonumber.
\end{eqnarray}
Such deformations of the $CS$ theory we reach here were probably ignored so
far because the solutions found for $\stackrel{\wedge }{A}_i$ have consisted of
cohomologically trivial variations (gauge transformations) of the same
particular solution of the SW map \cite{fidanza}. Nevertheless we have to
remark that the possible existence of nontrivial contributions had already
been anticipated in \cite{asakawa}.

The main point in the $\theta $ deformations in the commutative space
obtained in (\ref{37}) is that they are solely expressed in terms of the
curvature and derivatives. This indeed is not a specific feature of the
order that we have analysed. As we have shown in section $2$, the elements $%
\mathbb{A}_i^{(n)}\in \mathcal{H}(s)$ generate a covariant mapping, eq.(\ref{33})%
, in the SW map. Eq. (\ref{37}) is an example of this fact, as it is just a
mapping of the $CS$ action by $A_i\rightarrow A_i+\mathbb{A}_i^{(2)}$ up to
second order. So, we can say that the general form of the commutative action
after a SW map of the $NCCS$ taking into account all possible $\mathbb{A}%
_i^{(n)} $ is

\begin{equation}
\stackrel{\wedge }{S}_{NCCS}\longrightarrow S_{CS}+\frac{1}{2}\sum_{n=1}^\infty \int d^3x\epsilon ^{abc}%
\mathbb{A}_a^{(n)}\left[ F_{bc}+\sum_{m=1}^\infty \partial _b\mathbb{A}%
_c^{(m)}\right] ,  \label{38}
\end{equation}
and we see that the deformations of the $CS$ action are all given by monomials
constructed with the curvature and its derivatives. We can also assure that
they all are interaction terms. The only possible contribution to a kinetic
term would come from the first order (\ref{34}). But this first order term,
being a pure gauge, does not deform the $CS$ action.

In \cite{perturbing} actions as (\ref{38}) were studied. It was then shown
that even these non-power-counting interactions cannot change the
topological character of the $CS$ theory, at least, perturbatively. The
sensible point is that the kinetic topological action induces a definition
of the physical observables of the theory as link invariants, and these are
not perturbed by the interaction terms
(it is straightforward to generalize the argument in \cite{perturbing} to interactions with external parameters $\theta$).

We thus conclude that, in spite of the deformations appearing in the
action (\ref{38}), the SW map of the $NCCS$ leads to commutative actions
physically equivalent to the 3D Chern-Simons theory from the perturbative
point of view. This analysis, in this sense, complements the result of \cite
{gs}. This reasoning is also in agreement with the result of \cite{das},
where the authors showed that the tree level $CS$ coefficient is not
renormalized when $NCCS$ is quantized in the axial gauge.

\section{The General SW Map for NCMCS}
\setcounter{equation}{0}
Let us turn now to another example in $3D$ theories. The noncommutative
Maxwell-Chern-Simons model ($NCMCS$) has also been extensively studied \cite
{ghosh},\cite{cantcheff},\cite{dayi},\cite{mariz},\cite{harikumar}. The $%
NCMCS$ action is

\begin{equation}
\stackrel{\wedge }{S}_{NCMCS}=\int d^3x\left[ -\frac 14\stackrel{\wedge }{F}_{ij}*\stackrel{\wedge }{F}{}^{ij}+\frac m2\epsilon ^{ijk}\left( \stackrel{\wedge }{A}_i\partial _j\stackrel{\wedge }{A}_k-%
\frac{2i}3\stackrel{\wedge }{A}_i(\stackrel{\wedge }{A}_j*\stackrel{\wedge }{A}_k)\right) \right] ,
\label{39}
\end{equation}

\noindent where $m$ is the $3D$ noncommutative topological mass. The
particular solution of the $SW$ map leads to a complicated
non-power-counting commutative action, which has already been calculated up
to the second order in $\theta $ \cite{dayi}

\begin{eqnarray}
&&\stackrel{\wedge }{S}_{NCMCS}\stackrel{SW\text{ }map}{\longrightarrow }%
S_{MCS}-\frac{1}{2}\theta ^{ij}\int d^3x\left( F_{jk}F^{kl}F_{li}-\frac
14F_{ij}F^{kl}F_{kl}\right)  \nonumber \\
&&+\frac{1}{4}\theta ^{ij}\theta ^{kl}\int d^3x\left(
2F_{jk}F_{lm}F^{mn}F_{ni}+F_{jm}F_k^mF_l^nF_{ni}-F_{ij}F_{km}F^{mn}F_{nl}+%
\right.  \nonumber \\
&&\left. -\frac{1}{8}F_{ij}F_{kl}F_{mn}F^{mn}+\frac{1}{4}F_{jk}F_{li}F_{mn}F^{mn}\right) +O(\theta ^3).  \label{40}
\end{eqnarray}

Obviously, all these interaction terms come from the mapping of the
noncommutative Maxwell term in (\ref{39}), as the $NCCS$ term is not
transformed by the particular $SW$ map of (\ref{asw}) \cite{gs}. In \cite{ganor}
an argument was given for concluding that the interaction terms will always
depend only on the field strength $F$ at any order in $\theta $. This was
latter formally proved in \cite{berrino} so that

\begin{equation}
\stackrel{\wedge }{S}_{NCMCS}\longrightarrow S_{MCS}+\mathcal{L}(\theta ,F).
\label{41}
\end{equation}

In the previous section, we showed how the covariant contributions to the SW map
change the form of the commutative action coming from the $NCSS$ action. We
can ask here, in the $NCMCS$ case, what role the covariant contributions can
play.

The authors in \cite{harikumar} showed that a term of the form $\frac
1m\epsilon _{ijk}\theta ^{jk}F^2$ could be used to cancel part of the first
order in $\theta $ term in (\ref{40}), although other first order terms would
then be generated. This prompts us to generalize the idea a little bit
further and find the complete form of the covariant term that should be
added to (\ref{asw}) in order to cancel the first order terms in (\ref{40}). 
Indeed we found that 

\begin{equation}
\mathbb{A}_i^{(1)}=\frac{1}{2m}\epsilon _{ijk}\theta ^{lj}F^{kn}F_{nl}-\frac
1{8m}\epsilon _{ijk}\theta ^{jk}F^2 , \label{42}
\end{equation}
when substituted in the $NCCS$ sector of (\ref{39}), is able to cancel the first order contribution of (\ref{40}).
Notice that the terms in (\ref{42}) for $\mathbb{A}_i^{(1)}$ were not allowed in
the pure $CS$ case, eq. (\ref{34}), since they depend on the metric, which was
not part of the field and parameter content of the theory.
But when we substitute (\ref{42}) in the $NCM$ sector of (\ref{39}), we generate $\theta$ first order terms in $\frac{1}{m}$. These can again be cancelled by covariant terms in $\frac{1}{m^{2}}$ in the $NCCS$ sector, and we see that the complete $\mathbb{A}_i^{(1)}$ is in fact an infinite series in powers of $\frac{1}{m}$.
The same process will happen in each $\theta$ order of the covariant ambiguities $\mathbb{A}_i^{(n)}$.

The existence of such covariant mappings is not really a novelty. In \cite
{mapmcs}, the authors proved that covariant mappings can always be defined
in a way to reabsorb into the 3D free Maxwell-Chern-Simons action
interaction terms constructed only with the field strength $F$ and
derivatives, which is just the case of (\ref{41}). Further, as the covariant
mappings are part of the ambiguities which are allowed in the SW map, all
results of \cite{mapmcs} can be adapted to this present case.

With this intent, we begin by writing the generalization of (\ref{3.8a})
taking into account the covariant ambiguities of the SW map,

\begin{equation}
\delta _\theta ^{\prime }\stackrel{\wedge }{A}_i=-\frac 14\delta \theta
^{kl}\left\{ \stackrel{\wedge }{A}_k\stackrel{*}{,}\partial _l\stackrel{%
\wedge }{A}_i+\stackrel{\wedge }{F}_{li}\right\} +\delta \theta ^{kl}%
\stackrel{\wedge }{f}_{kli}\left( \stackrel{\wedge }{F},\stackrel{\wedge }{D}%
,\theta \right)  \label{43}
\end{equation}
where $\stackrel{\wedge }{f}_{kli}$ stands for all possible covariant terms constructed
with the curvatures $\stackrel{\wedge }{F}$, the Moyal covariant derivatives and the $%
\theta $ parameters. A generalization of this kind was first proposed in 
\cite{asakawa}, but only trivial terms (total derivatives) were considered
for $\stackrel{\wedge }{f}_{kli}$. Obviously, $\stackrel{\wedge }{f}_{kli}$ will transform covariantly
in the noncommutative sense:

\begin{equation}
s\stackrel{\wedge }{f}_{kli}=i\left[ \stackrel{\wedge }{C}\hspace{1mm}\stackrel{*}{,}%
\hspace{1mm}\stackrel{\wedge }{f}_{kli}\right] \text{ .}  \label{44}
\end{equation}
Using eqs.(\ref{43}) and (\ref{gs}), we get that after a general SW map the 
$NCCS$ action can only depend on $\theta $ through terms containing 
$\stackrel{\wedge }{f}_{kli} $,

\begin{equation}
\delta _\theta ^{\prime }\stackrel{\wedge }{S}_{NCCS}=\frac m2\epsilon^{ijk}\delta
\theta ^{mn}\int d^3x\stackrel{\wedge }{f}_{mni}\stackrel{\wedge }{F}_{jk}.
\label{45}
\end{equation}
\noindent But upon the $NCM$ action, both parts of $\delta _\theta ^{\prime }%
\stackrel{\wedge }{A}_i$ will contribute. First, we write the expression for 
$\delta _\theta ^{\prime }\stackrel{\wedge }{F}_{kl}$,

\begin{equation}
\delta_\theta ^{\prime }\stackrel{\wedge}{F}_{kl}=\frac{\delta \theta ^{mn}}{4}\left[
2\left\{ \stackrel{\wedge}{F}_{km}\stackrel{*}{,}\stackrel{\wedge}{F}_{ln}\right\} -\left\{ 
\stackrel{\wedge}{A}_m\stackrel{*}{,}\stackrel{\wedge}{D}_n\stackrel{\wedge}{F}_{kl}
+\partial _n\stackrel{\wedge}{F}_{kl}\right\} +4\left( \stackrel{\wedge}{D}_k\stackrel{\wedge}{f}_{mnl}
-\stackrel{\wedge}{D}_l\stackrel{\wedge}{f}_{mnk}\right) \right] ,  \label{46}
\end{equation}

\noindent and then 
\begin{eqnarray}
\delta _\theta ^{\prime }\stackrel{\wedge}{S}_{NCM} &=&-\frac 12\delta \theta
^{mn}\int d^3x\left[ \left( \frac 12\left\{ \stackrel{\wedge}{F}_{im}\stackrel{*}{,}\stackrel{\wedge}{F}_{jn}\right\} -\frac 14\left\{ \stackrel{\wedge}{A}_m\stackrel{*}{,}\stackrel{\wedge}{D}_n%
\stackrel{\wedge}{F}_{ij}+\partial _n\stackrel{\wedge}{F}_{ij}\right\} \right) \stackrel{\wedge}{F}{}^{ij}\right]  
\label{47} \nonumber \\ 
&&\hspace{1cm}-\frac 12\delta \theta
^{mn}\int d^3x\left[\left( \stackrel{\wedge}{D}_i\stackrel{\wedge}{f}_{mnj}-\stackrel{\wedge}{D}_j\stackrel{\wedge}{f}_{mni}\right) 
\stackrel{\wedge}{F}{}^{ij}\right] .
\end{eqnarray}

The sensible question is if it is possible to make

\begin{equation}
\delta _\theta ^{\prime }\stackrel{\wedge}{S}_{NCCS}+\delta _\theta ^{\prime }%
\stackrel{\wedge}{S}_{NCM}=0  \label{48}
\end{equation}
\noindent with a convenient choice of $\stackrel{\wedge}{f}_{ijk}$. The first step is to
write the second element in (\ref{47}) exclusively in terms of $\stackrel{\wedge}{F}%
_{ij}$. This can be achieved as

\begin{equation}  \label{49}
\delta\theta^{mn}\int d^3x\left\{\stackrel{\wedge} A_m\stackrel{*}{,}\stackrel{\wedge}
D_n\stackrel{\wedge} F_{ij}+\partial_n\stackrel{\wedge} F_{ij}\right\}\stackrel{\wedge} F{}^{ij}=\frac
12\delta\theta^{mn}\int d^3x\left\{\stackrel{\wedge} F_{mn}\stackrel{*}{,}\stackrel{\wedge}
F_{ij}\right\}\stackrel{\wedge} F{}^{ij}
\end{equation}

\noindent and (\ref{48}) leads to the equation

\begin{eqnarray}
\frac m2\epsilon ^{ijk}\delta \theta ^{mn}\int d^3x \stackrel{\wedge}{f}_{mni}\stackrel{\wedge}{F}_{jk}&=&\frac{\delta \theta ^{mn}}{2}\int d^3x\left( \frac 12\left\{ \stackrel{\wedge}{F}_{im}\stackrel{*}{,}\stackrel{\wedge}{F}_{jn}\right\} -\frac 18\left\{ \stackrel{\wedge}{F}_{mn}\stackrel{*}{,}\stackrel{\wedge}{F}_{ij}\right\} \right) \stackrel{\wedge}{F}{}^{ij} \nonumber \\ &+& \frac{\delta \theta ^{mn}}{2}\int d^3x \left(  2 \stackrel{\wedge}{f}_{mni}\stackrel{\wedge}{D}_j\stackrel{\wedge}{F}{}^{ij}\text{ }\right) .  \label{50}
\end{eqnarray}

This equation can be solved for $ \stackrel{\wedge}{f}_{ijk}$ as a series in powers of $%
\frac 1m$, 
\begin{equation}
\stackrel{\wedge}{f}_{ijk}=\sum_{r=1}^\infty \stackrel{\wedge}{f}_{ijk}^{\text{ }(r)}\text{ },
\label{51}
\end{equation}
where the upper index $r$ designates the power in $\frac 1m$. The first term
in the order $\frac 1m$ is

\begin{equation}
\stackrel{\wedge}{f}_{mni}^{\text{ }(1)}=\frac 1{4m}\epsilon _{nik}\left\{ \stackrel{\wedge}{F}_{jm}\stackrel{*}{,}\stackrel{\wedge}{F}{}^{jk}\right\} -\frac
1{16m}\epsilon _{mni}\left\{ \stackrel{\wedge}{F}_{jk}\stackrel{*}{,}\stackrel{\wedge}{F}{}^{jk}\right\}  \label{52}
\end{equation}
and the next orders can be obtained recursively from

\begin{equation}
\stackrel{\wedge}{f}_{mni}^{\text{ }(r+1)}=
-\frac 1m\epsilon_{ijk}\stackrel{\wedge}{D}^k\stackrel{\wedge}{f}_{mn}^{\text{ }(r)}{}^j . \label{53}
\end{equation}
This means that with the choice of this specific covariant polynomial in (\ref
{43}) we can assure the independence of the noncommutative
Maxwell-Chern-Simons theory on the $\theta $ parameter, eq.(\ref{48}), $ie$.

\begin{equation}
\delta _\theta ^{\prime }\stackrel{\wedge}{S}_{NCMCS}=0\text{ .}  \label{54}
\end{equation}

\section{Conclusion}
\setcounter{equation}{0}
In this paper we studied the effects of the ambiguities of the
Seiberg-Witten map on 3D noncommutative gauge theories. We showed how
covariant ambiguities added to the normally used SW map deform the commutative
Chern-Simons action by $%
\theta $ interaction terms. We also showed how choosing adequate
representatives of nontrivial (in the BRST sense) ambiguities the
noncommutative Maxwell-Chern-Simons action can be mapped into the
commutative version without the interaction terms which are usually
associated to the SW mapping of Maxwell type actions.

Although these two results apparently point out to different directions,
they are, in fact, the same development seen from different points of view.
In both cases, what we have shown is that among all possible SW maps
classified by BRST cohomology, we can find one element which cancels the $%
\theta $ contributions. It thus erases the memory of these theories from their
noncommutative origin. They are mapped into renormalizable, and in the
abelian cases we treat, even free, commutative theories. The only difference
between both cases lies in the need for a nontrivial BRST element in the $%
NCMCS$ case, whereas the $NCCS$ case requires that only trivial BRST terms
should be added to the particular solution of the SW map in order to reach
the commutative pure $CS$ theory. 

Indeed, both cases have more in common. As it was shown in \cite{mapmcs},
what allows for the reabsorption of interactions made of covariant elements
(such as curvatures and their covariant derivatives) by non-linear local
gauge field redefinitions is the presence of the Chern-Simons term in the
action (in \cite{ozemar e rubens} the pure Maxwell theory
is mapped to pure Chern-Simons theory but that mapping is not a series of local
terms). This was crucial in finding the solution to eq. (\ref{50}), which
led to eq. (\ref{54}). We believe that this is a general feature of
noncommutative actions with Schwarz type topological sectors (we will be
reporting soon on 4D noncommutative BF theories \cite{underdevelopment}).We
also believe that these theories will be well-behaved upon quantization, as
long as gauge fixing conditions and all quantization procedures can be
translated from the commutative to the noncommutative space (this has been
shown for the pure $NCCS$ theory by explicit calculations in the
noncommutative space \cite{das}), although the renormalizability of pure $%
NCBF$ models has been questioned in \cite{blasi}. Unfortunately, the same
argument indicates that purely geometrical theories, as for instance the
pure noncommutative Maxwell theory, seem to have unavoidably the companion
of power counting nonrenormalizable $\theta $ interactions in their
commutative versions after the SW map. The already found
nonrenormalizability of $NCQED$ \cite{WUL} is an evidence of this fact. The
ambiguities of the SW map seem to be of no hope in these cases.

\section*{Acknowledgments}

The Conselho Nacional de Desenvolvimento Cient\'{i}fico e Tecnol\'{o}gico
(CNPq-Brazil), the Funda{\c{c}}{\~{a}}o de Amparo \`a Pesquisa do Estado do Rio
de Janeiro (Faperj), the SR2-UERJ, the Funda\c c\~ao de Apoio \`a Educa\c c\~ao,
Pesquisa e Desenvolvimento Tecnol\'ogico (FUNCEFETES) and the Centro Universit\'ario
de Vila Velha (UVV) are acknowledged for financial support.

\end{document}